# Developing a Responsible AI Framework for Healthcare in Low Resource Countries: A Case Study in Nepal and Ghana

Hari Krishna Neupane[1], Bhupesh Kumar Mishra[2]
[1]PhD Researcher, University of Hull, Cottingham Rd, Hull HU6 7RX, Hull
[2]Lecturer in AI and Data Science, University of Hull, Cottingham Rd, Hull HU6 7RX, Hull
*Correspondence Email: h.neupane-2022@hull.ac.uk

**Abstract**

The integration of Artificial Intelligence (AI) into healthcare systems in low-resource settings, such as Nepal and Ghana, presents transformative opportunities to improve personalized patient care, optimize resources, and address medical professional shortages. This paper presents a survey-based evaluation and insights from Nepal and Ghana, highlighting major obstacles such as data privacy, reliability, and trust issues. Quantitative and qualitative field studies reveal critical metrics, including 85% of respondents identifying ethical oversight as a key concern, and 72% emphasizing the need for localized governance structures. Building on these findings, we propose a draft Responsible AI (RAI) Framework tailored to resource-constrained environments in these countries. Key elements of the framework include ethical guidelines, regulatory compliance mechanisms, and contextual validation approaches to mitigate bias and ensure equitable healthcare outcomes.

**Key Words:** Artificial Intelligence, Healthcare, Responsible AI, Low-Resource Settings, Nepal, Ghana, Ethical AI, Low and Middle Income countries (LMICs).

## 1.    Introduction

The adoption of AI in healthcare has brought substantial advancements, improving diagnostic accuracy, treatment personalization, and administrative efficiency. While developed countries have successfully incorporated AI into their healthcare systems, LMICs such as Nepal and Ghana face unique challenges, including a lack of digital infrastructure, policy gaps, and ethical concerns. Despite its potential, AI implementation remains hindered by privacy, liability, and trust issues. This paper aims to evaluate AI healthcare initiatives in Nepal and Ghana, analyse risks and benefits, and propose a Responsible AI Framework draft, tailored to low-resource settings.

The examination of digital health initiatives in Nepal and Ghana has revealed that AI is becoming an integral component of healthcare systems in these nations. However, the widespread use of AI remains largely unregulated, lacking the necessary oversight and ethical safeguards to ensure its responsible application. This absence of a structured AI framework creates uncertainty among patients, regulators, and AI developers, raising concerns regarding safety, fairness, and accountability (Floridi et al., 2020). Without appropriate governance, AI could exacerbate existing healthcare disparities, reinforce biases, and introduce vulnerabilities that compromise patient outcomes (Obermeyer et al., 2019). Hence, there is an urgent need to develop a RAI Framework that provides clear ethical guidelines, regulatory compliance, and localized validation mechanisms to ensure AI adoption aligns with the unique needs of low-resource healthcare settings (WHO, 2021).

## 1.1 Research Questions

The study aims to address the following research questions, ensuring alignment with the methodology and findings:

- What are the benefits and risks associated with AI healthcare initiatives in low-resource settings such as Nepal and Ghana?
- What are the key challenges hindering the integration of AI in healthcare within these low-resource environments?
- How can a Responsible AI Framework be developed and validated to address the unique socio-economic and technological constraints of low-resource healthcare systems?

## 1.2 Research Methodology

This study employs a mixed-methods approach, combining qualitative and quantitative fieldwork conducted in Nepal and Ghana. Key methodological elements include:

- Literature review: Combining literature reviews of existing AI frameworks and policies, field studies around developed nations, Nepal and Ghana.
- Sampling: Surveys were distributed to healthcare practitioners, policymakers, and AI developers, with a total of 85 respondents across both nations.
- Survey Design: Questions focused on ethical concerns, trust issues, and governance gaps related to AI deployment.
- Data Collection: Field studies were conducted in urban and rural healthcare facilities, ensuring diverse representation.
- Ethical Approvals: The study adhered to ethical standards approved by the University of Hull, UK.

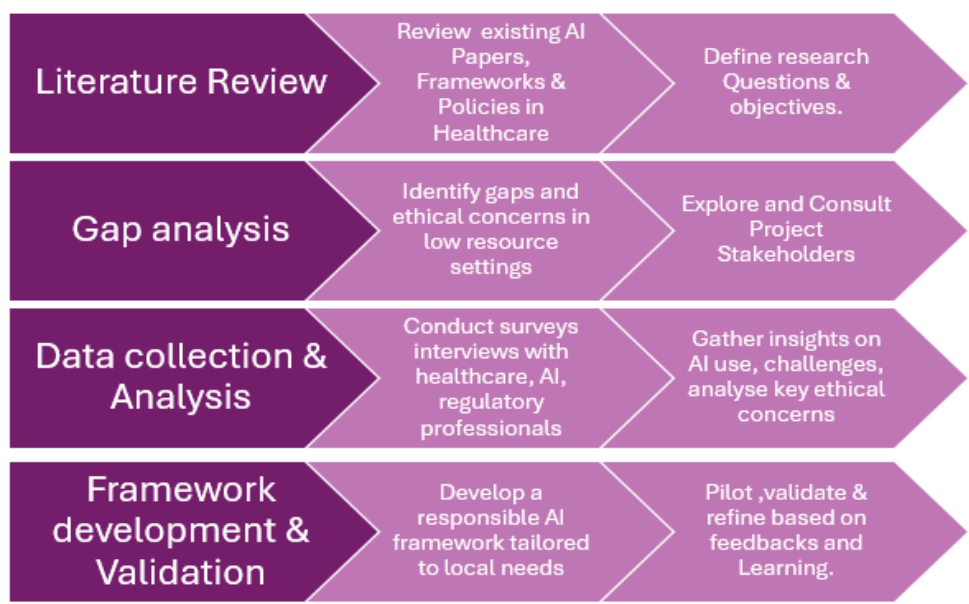

*Figure 1: Conceptual flow diagram of adopted methodology.*

## 2. Literature Review

### 2.1 Existing AI Frameworks in Healthcare

Artificial Intelligence has gained significant traction in healthcare, leading to the development of various ethical and regulatory frameworks to ensure its responsible deployment. The WHO AI4Health Initiative provides guidelines for AI governance in healthcare, emphasizing transparency, safety, and data privacy (WHO, 2021). Similarly, the European AI Ethics Guidelines advocate for human oversight, fairness, and accountability in AI-driven healthcare solutions (European Commission, 2019). In the US, the AI Bill of Rights was introduced to protect patients from algorithmic biases and ensure that AI systems remain interpretable and safe (WhiteHouse, 2022). Despite these initiatives, there is a notable lack of contextual adaptation for LMICs, where data scarcity, infrastructure gaps, and ethical concerns present significant barriers (Rahman et al., 2021).

### 2.2 AI in the UK's NHS

The National Health Service (NHS) has actively integrated AI in diagnostics, resource management, and predictive analytics to enhance patient care (Topol, 2019). AI-driven imaging tools have improved the early detection of diseases such as cancer, significantly reducing diagnostic errors (Rajpurkar et al., 2018). However, ethical concerns persist, particularly regarding data privacy, algorithmic biases, and regulatory challenges. Studies highlight that while AI adoption enhances efficiency, issues such as transparency and trust among healthcare professionals remain (Holdsworth & Zaghloul, 2022). Recent NHS policies now emphasize the need for explainability and regulatory oversight to ensure that AI complements human decision-making rather than replacing it (UK Government, 2022).

### 2.3 AI Policies in Developing Nations

AI policies in developing nations remain fragmented and underdeveloped. Countries like India and South Africa have introduced AI ethics guidelines focusing on data privacy, inclusivity, and fairness (NITI Aayog, 2020). India's Responsible AI for Healthcare framework highlights the importance of ethical AI deployment while ensuring compliance with global standards such as the GDPR and WHO's AI guidelines (NITI Aayog, 2020). In contrast, Ghana and Nepal lack formal AI policies, resulting in inconsistent AI adoption and regulatory uncertainty (UNESCO, 2023). Studies suggest that developing nations need tailored AI governance strategies that consider socio-economic factors, healthcare disparities, and technological constraints (Vayena et al., 2018).

A systematic review by Obermeyer et al. (2019) identified that most AI-driven healthcare solutions in LMICs fail due to data bias and inadequate validation of local populations. Similarly, research by Ting et al. (2020) emphasizes the need for clinical trials and real-world validation before AI tools are deployed in resource-limited healthcare settings. Further, studies by Panch et al. (2019) and Beauchamp & Childress (2019) stress the importance of ethical considerations, highlighting the need for frameworks that ensure fairness, transparency, and accountability in AI-driven healthcare systems. These findings collectively indicate that while AI presents immense potential, its deployment in LMICs must be accompanied by robust policies, regulatory oversight, and ethical considerations to maximize its benefits while mitigating risks.

## 3. Challenges in AI Implementation in Developing Countries

### 3.1 Data Privacy and Security

One of the primary concerns in AI deployment in healthcare is data privacy and security. Many developing countries lack robust data governance frameworks, increasing the risk of patient data breaches due to inadequate cybersecurity measures. Studies indicate that without proper regulatory oversight, AI-driven healthcare applications may inadvertently expose sensitive patient information, leading to ethical and legal challenges (Ting et al., 2020). The absence of standardized encryption protocols and weak cybersecurity measures exacerbates vulnerabilities, making AI systems prime targets for cyber threats (Rahman et al., 2021). Ensuring compliance with international data protection laws, such as the General Data Protection Regulation (GDPR), remains a significant challenge in LMICs (Vayena et al., 2018).

### 3.2 Algorithmic Bias and Trust Issues

AI models trained on Western datasets often fail to generalize in LMICs healthcare settings, leading to skewed results and potential misdiagnoses. The lack of diverse, region-specific datasets results in biased predictions that disproportionately affect underserved populations (Obermeyer et al., 2019). Studies highlight that healthcare professionals in LMICs exhibit skepticism toward AI recommendations due to a lack of transparency and explainability in AI decision-making processes (Panch et al., 2019). To build trust in AI systems, healthcare institutions must ensure contextual validation and algorithmic fairness, reduce biases and improve AI interpretability (Beauchamp & Childress, 2019).

### 3.3 Infrastructure and Workforce Limitations

The successful deployment of AI in healthcare requires robust digital infrastructure and a well-trained workforce. However, many LMICs face challenges such as limited internet connectivity, outdated hospital IT systems, and a shortage of AI-trained medical professionals and data scientists (WHO, 2021). A lack of continuous professional development programs for clinicians further hampers AI adoption, as many healthcare workers are unfamiliar with AI-based decision support tools (Ting et al., 2020). Moreover, inadequate funding for AI research and implementation restricts the scalability of AI-driven solutions in LMICs. Addressing these challenges requires investment in digital infrastructure, workforce training, and cross-sector collaboration to bridge the AI readiness gap in healthcare systems (Rahman et al., 2021).

## 4. Understanding Healthcare Ecosystem in Nepal and Ghana: Field Study and Survey

### 4.1 Healthcare Ecosystem in Nepal

Nepal's healthcare system has made significant progress in recent years, yet it continues to face challenges in adopting digital health technologies and AI integration. Through discussions with digital health experts, healthcare professionals, and policy regulators, it has been observed that Nepal is gradually adopting Electronic Health Record (EHR) systems and Hospital Management Information Systems (HMIS) to improve patient data management. However, these systems remain fragmented due to a lack of interoperability across public and private healthcare sectors. The inconsistency in data standardization poses a critical challenge, as healthcare providers struggle to maintain well-defined patient records. Furthermore, regulatory gaps and weak data privacy policies raise concerns about ethical AI adoption in healthcare. Although there is an increasing interest in AI-driven healthcare solutions, the country lacks a comprehensive framework to guide responsible AI deployment. There is an urgent need for AI governance policies, role-based data security, and collaboration between policymakers, AI developers, and healthcare institutions to ensure ethical, effective, and secure AI integration in Nepal.

### 4.2  Healthcare Ecosystem in Ghana

Ghana's healthcare system is supported by multiple digital health initiatives, including the District Health Information Management System (DHIMS2) and Lightwave Health Information Management System (LHIMS), which aid in patient records management and health service delivery. Additionally, the National Health Insurance Authority (NHIA) system plays a crucial role in claims processing and healthcare financing. The E-Pharmacy project is another government initiative aimed at providing telehealth and remote prescription services, enhancing accessibility to medical consultations. However, Ghana's healthcare digitalization faces substantial hurdles, such as fragmented health data systems, connectivity issues, and inadequate AI policy frameworks. Stakeholder discussions indicate a pressing need for AI-driven data analytics, predictive modelling for disease surveillance, and AI-based decision-support systems. Survey findings reveal that although healthcare professionals acknowledge AI's potential in improving patient care, they also express concerns about bias in AI models, lack of transparency, and low AI literacy among medical practitioners. Therefore, structured AI training programs, regulatory oversight, and strategic public-private partnerships are essential to facilitate responsible AI integration in Ghana's healthcare sector.

### 4.3  Data Collection and Analysis: Survey

We gathered important insights and feedback from an initial survey conducted among 85 respondents, including healthcare professionals, IT experts, and regulatory personnel in Nepal and Ghana. The survey aimed to assess AI awareness, usage, perceived risks, and local mitigation strategies. Below is a statistical summary of the key findings, followed by a critical analysis of how these findings influence the proposed Responsible AI (RAI) framework.

## 5.  Survey Findings and Statistical Summaries:

### 5.1  Professional Diversity: Work Domain and Job Titles

The majority of respondents (53%) were healthcare professionals, while IT experts and regulatory personnel constituted 15% and academia 12%, respectively. This distribution highlights the multidisciplinary involvement required for AI adoption in healthcare (Figure 2).

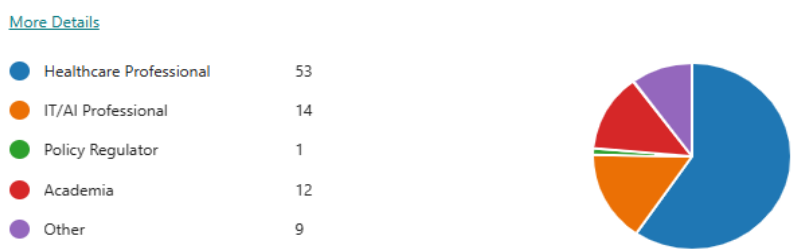

*Figure 2: Professional diversity of the survey participants.*

### 5.2    Perception of AI's Impact on Patient Care

The majority opinion is that AI integration will improve the quality of patient care, as presented in Figure 3.

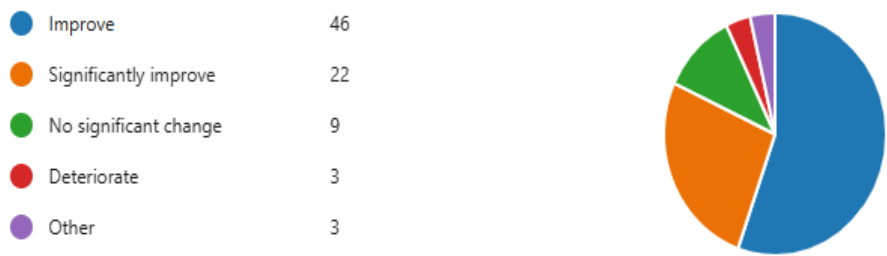

*Figure 3: Participants' opinions on the impact of AI integration into healthcare.*

### 5.3    Necessity of a Responsible AI Framework

A strong consensus emerged on the importance of enforcing a Responsible AI framework. Key recommendations included:

- Strict guidelines and regulations (61%)

- Moderate regulation (15%)

- No regulation (0%)

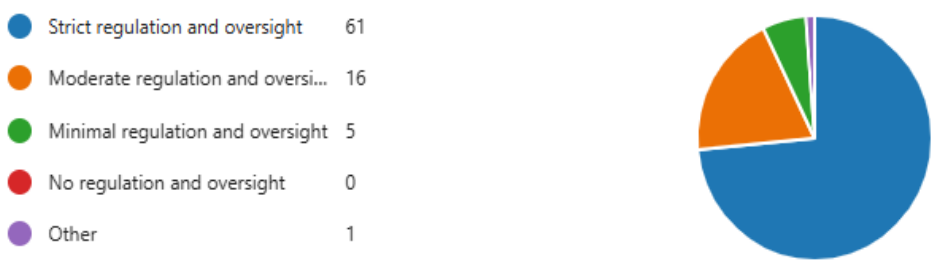

*Figure 4: Participants' opinions on the necessity of a Responsible AI Framework.*

### 5.4    Challenges Identified

Low AI literacy among healthcare workers (65%), lack of robust infrastructure (58%), and limited funding for AI research (52%) were the most frequently cited challenges.

### 5.5    Critical Analysis

The survey findings emphasize the need for a localized and context-sensitive approach to integrating AI into healthcare in developing countries, such as Nepal and Ghana. Below is an analysis of how the results shape the proposed RAI framework:

## 6. Implementation Strategies & Draft Framework

Effective Implementation of Responsible AI in the healthcare sector of developing countries has got many challenges and needs to be done with careful consideration. To integrate AI efficiently into healthcare in developing nations, digital capacity building is a fundamental necessity. Training healthcare workers in AI literacy and ethical guidelines is a critical first step, ensuring that professionals understand how to interpret AI-generated insights and integrate them into clinical workflows (Ting et al., 2020). Educational initiatives should focus on developing digital competencies, enabling clinicians to critically assess AI recommendations and intervene when necessary (Beauchamp & Childress, 2019). Public-private partnerships (PPPs) can play a pivotal role in fostering collaborations between governments, technology firms, and healthcare providers. PPPs facilitate resource sharing, reducing the financial burden on governments while promoting AI research and innovation in LMICs (Rahman et al., 2021). These partnerships can also aid in creating AI solutions that are tailored to local needs, ensuring that technology adoption aligns with regional healthcare priorities (NITI Aayog, 2020).

Conducting pilot studies is necessary to test AI deployments on a small scale before national implementation. Pilot programs provide empirical data on AI performance in real-world settings, highlighting areas for improvement before large-scale rollout (Panch et al., 2019). These trials should assess the accuracy, fairness, and usability of AI applications, ensuring that models function optimally across diverse patient populations (WHO, 2021). Furthermore, continuous monitoring and evaluation mechanisms must be established to ensure AI-driven healthcare applications remain effective and aligned with ethical standards. Developing real-time feedback loops, where AI-generated outcomes are routinely assessed and refined, will help maintain system reliability and mitigate unintended biases (Obermeyer et al., 2019). By prioritizing ongoing assessment, LMICs can ensure that AI remains a tool for enhancing, rather than compromising, healthcare quality and accessibility.

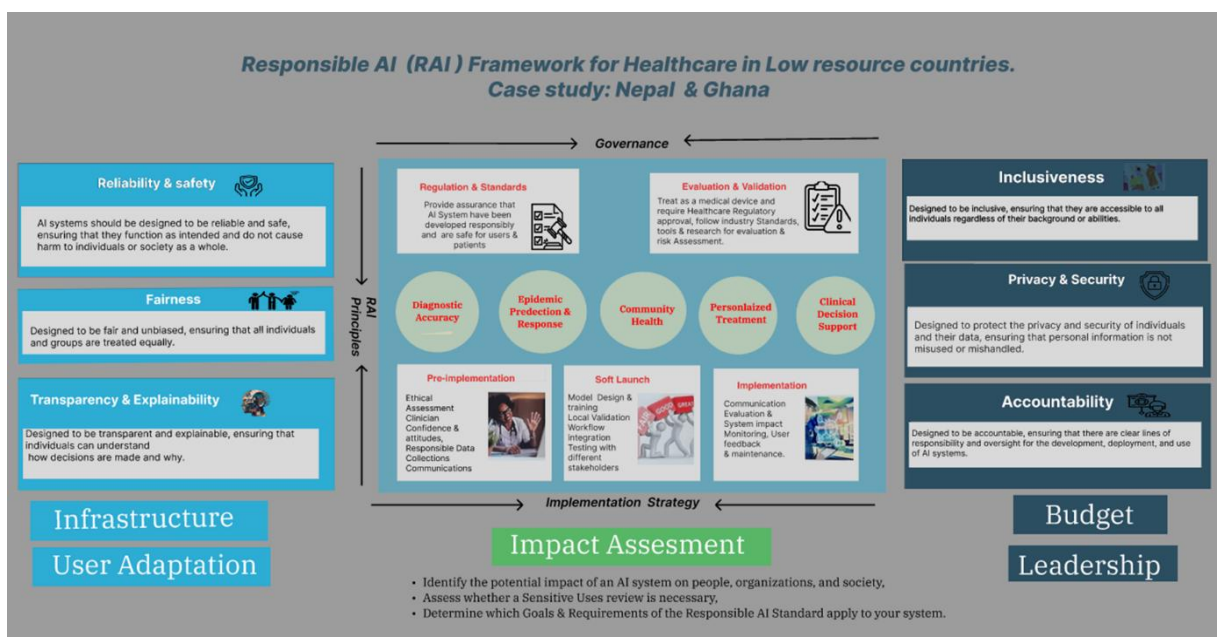

*Figure 5: Draft Responsible AI Framework for Low-Resource Countries.*

The insights gained from this study reinforce the necessity of establishing the RAI Framework, as presented in Figure 5 that fosters trust, compliance, and equitable access to AI-driven healthcare innovations. The draft RAI Framework is designed to ensure ethical and effective AI implementation. It is built around key RAI principles: Reliability & Safety, Fairness, Transparency & Explainability, Inclusiveness, Privacy & Security, and Accountability.

The framework integrates governance through regulatory standards and validation processes while ensuring smooth Implementation via ethical assessment, stakeholder engagement, and monitoring systems. It follows a structured approach—beginning with a literature review to identify gaps, followed by issue identification through surveys and interviews. The framework is then tailored to local needs, piloted, and refined based on stakeholder feedback. A strong emphasis is placed on infrastructure, user adaptation, budget, and leadership to ensure a sustainable and impactful AI-driven healthcare system. This is a draft framework under review from different stakeholders and has not been tested and validated yet. After having a through review this will be tested and validated.

## 7.  Conclusion and Future Work

In summary, this study examined the benefits and risks of AI in healthcare for low-resource settings, identifying key challenges such as limited digital literacy, infrastructure and lack of leadership commitment. Addressing these, we are trying to develop a Responsible AI Framework tailored to local socio-economic and technological realities and outlined steps for its validation. These findings directly address the research questions and demonstrate how ethical AI can improve healthcare in places like Nepal and Ghana, supporting equitable, effective solutions.

Field studies highlighted the importance of localised governance, collaboration with stakeholders, and strengthening AI literacy among healthcare professionals and policymakers. Addressing infrastructural gaps and establishing robust regulatory mechanisms are vital to ensure transparency, accountability, and equitable access. By adopting a Responsible AI framework, healthcare systems in these regions can reduce disparities and promote equitable, patient-centric outcomes.

Future work will focus on the need to strengthen AI literacy and awareness programs among healthcare professionals, policymakers, and AI practitioners, equipping them with the necessary knowledge to integrate AI effectively into healthcare systems (Ting et al., 2020). Additionally, field assessments will be conducted to evaluate the digital readiness of healthcare facilities, identifying infrastructural gaps and capacity-building opportunities (Rahman et al., 2021). Collaboration with health ministries, AI researchers, and regulatory bodies will be prioritized to develop context-specific AI policies that align with national healthcare objectives (Beauchamp & Childress, 2019). Furthermore, advocacy efforts will aim to establish robust legal and regulatory mechanisms, ensuring that AI implementation remains transparent, accountable, and ethically aligned with patient-centric healthcare. Through a multi-stakeholder approach, draft RAI framework will be validated with some use cases and harnessed to enhance healthcare delivery, reduce disparities, and promote long-term sustainability in LMIC healthcare ecosystems. Future research should also focus on pilot testing the framework and expanding its applicability to other LMICs and as per the local contexts of developing countries.